\documentclass[11pt]{article}
\usepackage{moriond,epsfig,amsmath}
\usepackage{relsize}
\RequirePackage{xspace}

\def\be{\begin{equation}}
\def\ee{\end{equation}}
\def\bea{\begin{eqnarray}}
\def\eea{\end{eqnarray}}

\newcommand{\btosg}{$b\rightarrow s\gamma$}
\newcommand{\btoksll}{$B\rightarrow K^*\ell^+\ell^-$}
\newcommand{\btoktl}{\ensuremath{B \to K \tau \ell}}
\newcommand{\btoptl}{\ensuremath{B \to \pi \tau \ell}}
\def\nut        {\ensuremath{\nu_\tau}\xspace}
\def\KS    {\ensuremath{K^0_{\scriptscriptstyle S}}\xspace}
\def\KL    {\ensuremath{K^0_{\scriptscriptstyle L}}\xspace}
\def\babar{\mbox{\slshape B\kern-0.1em{\smaller A}\kern-0.1em
    B\kern-0.1em{\smaller A\kern-0.2em R}}}
\def\Bbar    {\kern 0.18em\overline{\kern -0.18em B}{}\xspace}
\newcommand{\gevcc}{\ensuremath{{\mathrm{\,Ge\kern -0.1em V\!/}c^2}}\xspace}
\newcommand{\mevcc}{\ensuremath{{\mathrm{\,Me\kern -0.1em V\!/}c^2}}\xspace}
\def\mes       {\mbox{$m_{\rm ES}$}\xspace}
\def\jpsi     {\ensuremath{{J\mskip -3mu/\mskip -2mu\psi\mskip 2mu}}\xspace}
\def\psitwos  {\ensuremath{\psi{(2S)}}\xspace}
\def\nunub      {\ensuremath{\nu{\overline{\nu}}}\xspace}
\def\nub        {\ensuremath{\overline{\nu}}\xspace}
\def\nut        {\ensuremath{\nu_\tau}\xspace}
\def\nutb       {\ensuremath{\nub_\tau}\xspace}
\newcommand{\tautoenunu}{\ensuremath{\tau \to e \nunub}\xspace}
\newcommand{\tautomununu}{\ensuremath{\tau \to \mu \nunub}\xspace}
\newcommand{\tautopinu}{\ensuremath{\tau \to (n\pi^0) \pi \nu}\xspace}
\def\Dbar    {\kern 0.2em\overline{\kern -0.2em D}{}\xspace}
\def\taupksb     {\ensuremath{\tau^+\rightarrow\pi^+\KS\ \nutb}\xspace }
\def\taupks      {\ensuremath{\tau^-\rightarrow\pi^-\KS\,\nut}\xspace }
\def\asy         {\ensuremath{A_Q}\xspace }

\begin{document}
\vspace*{4cm}
\title{A SELECTION OF RECENT RESULTS FROM THE BABAR EXPERIMENT}

\author{V. Poireau, on behalf of the \babar\ collaboration}

\address{Laboratoire d'Annecy-le-Vieux de Physique des Particules (LAPP), \\Universit\'e de Savoie, CNRS/IN2P3}

\maketitle\abstracts{We present recent results from the \babar\ collaboration in several areas of research. These include searches for new physics via measurements of radiative-penguin $B$ decays, lepton-number and lepton-flavor violations in $B$ decays, and $CP$ violation in tau lepton decays.}

\section{Introduction}

Until 2008, the \babar\ experiment recorded $e^+e^-$ collisions at the $\Upsilon(4S)$ resonance with $471 \times 10^6$ $B\Bbar$ pairs produced (corresponding to an integrated luminosity of 429~fb$^{-1}$). Four years after the end of the data-taking period, \babar\ is still producing many new and interesting results. Despite the start of the LHCb experiment at CERN, \babar\ is still competitive, especially for channels involving neutral particles (such as photons, $\pi^0$ or \KS), tau leptons, and neutrinos.

We present here a selection of recent results from the \babar\ experiment. We show exclusive measurements of the \btosg\ transition rate as well as a study of the photon energy spectrum. Then we present an analysis of the angular distributions in the decay \btoksll. We detail a search for lepton-number violating processes in $B^+ \to h^- \ell^+ \ell^+$ decays, as well as a search for the decay $B^\pm \to h^\pm \tau \ell$. Finally, a search for $CP$ violation in the $\tau^-\to\pi^-\KS\left(\geq 0\pi^0\right)\nut$ channel is presented.

\section{Exclusive measurements of \btosg\ transition rate and photon energy spectrum}

As it is well known, flavor changing neutral currents (FCNC), such as \btosg, are forbidden at tree level in the standard model (SM). However, FCNC of this type are predicted to occur at loop level with the following rate~\cite{ref:misiak}: $\mathcal{B}(\Bbar\rightarrow X_{s}\gamma)=(3.15\pm0.23)\times 10^{-4}$,
for a minimum photon energy $E_\gamma>1.6$ GeV measured in the $B$ meson rest frame, and where $X_s$ is the final state of the $s$ quark hadronic system. The world average experimental value~\cite{ref:hfag} is measured at
$\mathcal{B}(\Bbar\rightarrow X_{s}\gamma)=(3.55\pm0.25\pm0.09)\times 10^{-4}$,
for $E_\gamma>1.6$ GeV, and where the second uncertainty is due to the extrapolation from the experimental photon energy (between 1.7 and 2.0 GeV depending on the experiments) to 1.6 GeV. The calculation is performed at next-to-next-leading order in the perturbative term, with the first order radiative penguin diagram for the \btosg\ transition having a $W$ boson and a $t$, $c$, or $u$ quark in the loop. Comparing the experimental and predicted values allows a precision test of the SM. Particles from new physics could enter in the loop, and would affect this transition rate. Furthermore, the photon energy spectrum from this reaction gives insight into the momentum distribution function of the $b$ quark inside the $B$ meson, and helps to constrain the uncertainty on the Cabibbo-Kobayashi-Maskawa matrix element $V_{ub}$.

In this \babar\ analysis~\cite{ref:babarbtosg}, we use a ``sum of exclusives'' approach, where we reconstruct 38 different $X_s$ final states (listed in Table~\ref{tab:btosg}). By fully reconstructing $X_s$, we obtain the energy of the transition photon in the $B$ rest frame with $E_{\gamma}^{B} = \frac{m_{B}^{2}-m_{X_{s}}^{2}}{2m_{B}}$, where $m_B$ is the $B$ mass, and $m_{X_{s}}$ is the invariant mass of the $X_s$ system. We use a range of $0.6 < m_{X_{s}} < 2.8$ \gevcc, which corresponds to a photon energy range of $1.9 < E_\gamma < 2.61$ GeV. In the analysis, this range is divided in 18 different $m_{X_{s}}$ regions.

Two types of signal Monte Carlo (MC) events are generated: one for the $K^*(892)$ region (corresponding to $m_{X_{s}}<1.1$ \gevcc) and one for the inclusive region (corresponding to $1.1 < m_{X_{s}} < 2.8$ \gevcc). A flat photon spectrum level is used in the inclusive region at the generation level, which allows us to reweight to match whichever spectrum model we choose.

\begin{figure}[htb]
\begin{center}
\includegraphics[width=7.5cm]{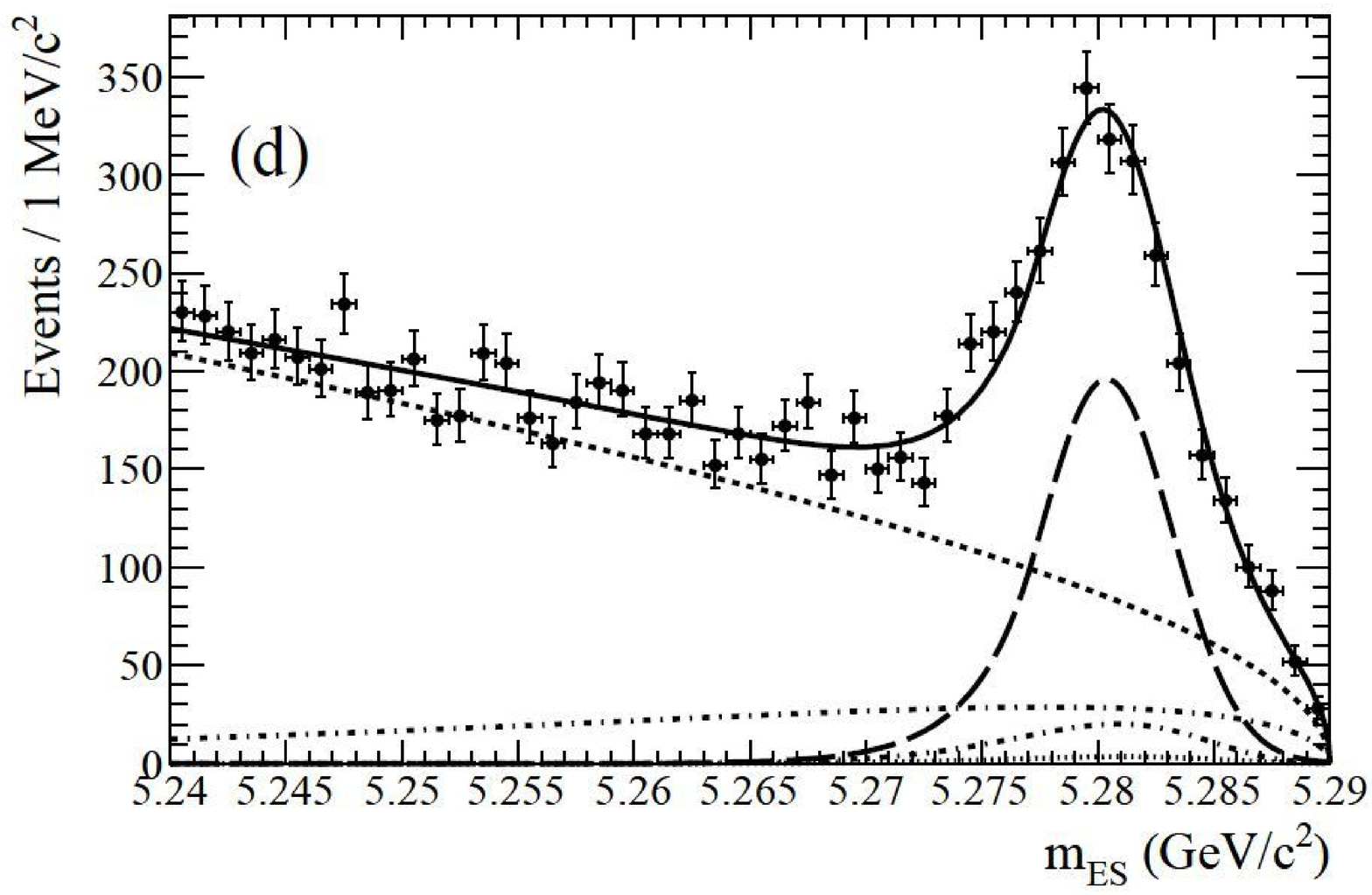}
\includegraphics[width=7.5cm]{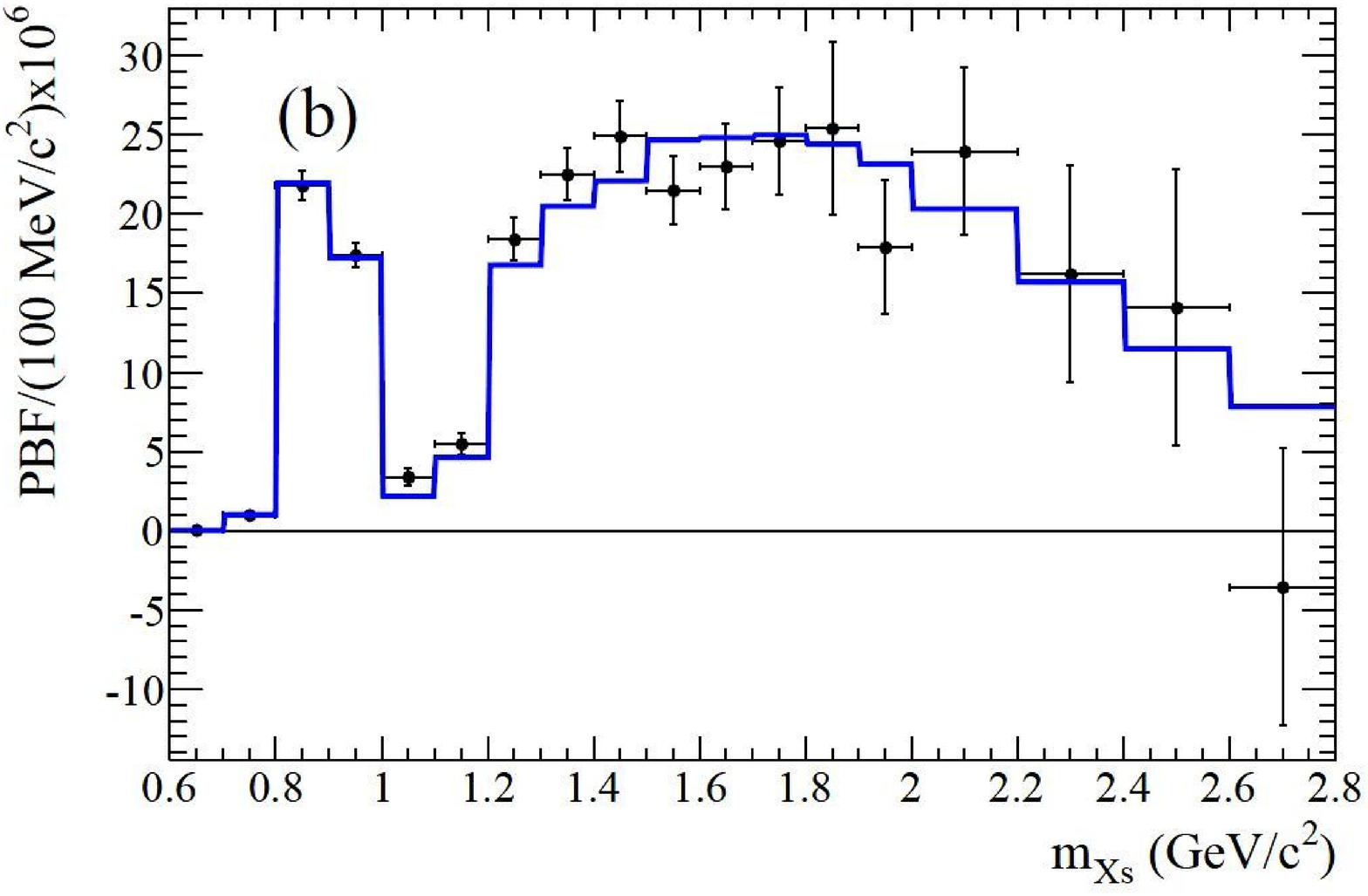}
\caption{Left: example of \mes\ fit for the region $1.4 < m_{X_{s}} < 1.5 \gevcc$. Points with statistical errors are the data, the solid line represents the total fit, the thick-dashed line represents the signal, and the other lines show background contributions. Right: partial branching fractions in each $m_{X_{s}}$ region (dots) and fit of the spectrum using the kinetic model (solid line).}
\label{fig:btosg}
\end{center}
\end{figure}

We use three classifiers (based on random forest classifiers) to reject the background, each optimized in four different $m_{X_{s}}$ regions. These three classifiers help respectively to choose the best $B$ candidate, to veto photons coming from a neutral pion, and to reject the continuum background. The signal yield is extracted in each $m_{X_{s}}$ region from a fit to the beam-energy substituted mass, $m_{ES}\equiv\sqrt{(\sqrt{s}/2)^{2}-(p_{B}^{*})^{2}}$, where $p_B^*$ is the momentum of the reconstructed $B$ meson in the center-of-mass. An example of such a fit is shown in Fig.~\ref{fig:btosg}.

The fragmentation of the hadronic system is modeled with \texttt{JETSET} with a phase-space hadronization model. We observe some differences between fragmentation in the MC and data samples. To correct this, we group the final states by topology and correct the signal contribution in the MC to better reflect the data. Furthermore, several quark hadronization models are tested and the observed differences are included in the systematic uncertainties.

The result of the measurements of the partial branching fractions in each $m_{X_{s}}$ region is shown in Fig.~\ref{fig:btosg}. We fit this spectrum with two models: the kinetic model and the shape-function model~\cite{ref:models}. The result is shown in Fig.~\ref{fig:btosg} for the kinetic model. We have also extracted the parameters from the heavy quark effective theory (HQET) for these two models, obtaining values compatible with the world average. The mean and variance of the photon energy spectrum have also been extracted, which can be used to constrain different kinds of models.
Summing the 18 bins, we find the total branching fraction for $E_\gamma>1.9$ GeV to be $\mathcal{B}(\Bbar\rightarrow X_{s}\gamma)=(3.29\pm0.19\pm0.48)\times 10^{-4}$, where the first uncertainty is statistical and the second is systematic. 

\begin{table}[htp]
\begin{small}
\begin{center}
\caption{\label{tab:btosg} The 38 final states of the $B$ meson reconstructed in this analysis.}
\begin{tabular}{|l|l|l|l|}
\hline
\multicolumn{4}{|c|}{Final states} \\
\hline
$ K_{S}\pi^{+}\gamma$ & $ K_{S}\pi^{+}\pi^{-}\pi^{+}\gamma$ & $ K^{+}\pi^{+}\pi^{-}\pi^{-}\pi^{0}\gamma$ & $ K_{S}\eta\pi^{+}\pi^{-}\gamma$ \\
$ K^{+}\pi^{0}\gamma$ & $ K^{+}\pi^{+}\pi^{-}\pi^{0}\gamma$	& $ K_{S}\pi^{+}\pi^{-}\pi^{0}\pi^{0}\gamma$ & $ K^{+}\eta\pi^{-}\pi^{0}\gamma$\\
$ K^{+}\pi^{-}\gamma$ & $ K_{S}\pi^{+}\pi^{0}\pi^{0}\gamma$ & $ K^{+}\eta\gamma$ & $ K^{+}K^{-}K^{+}\gamma$\\
$ K_{S}\pi^{0}\gamma$ & $ K^{+}\pi^{+}\pi^{-}\pi^{-}\gamma$	& $ K_{S}\eta\gamma$ & $ K^{+}K^{-}K_{S}\gamma$\\
$ K^{+}\pi^{+}\pi^{-}\gamma$ & $ K_{S}\pi^{0}\pi^{+}\pi^{-}\gamma$ & $ K_{S}\eta\pi^{+}\gamma$	& $ K^{+}K^{-}K_{S}\pi^{+}\gamma$\\
$ K_{S}\pi^{+}\pi^{0}\gamma$ & $ K^{+}\pi^{-}\pi^{0}\pi^{0}\gamma$ & $ K^{+}\eta\pi^{0}\gamma$	& $ K^{+}K^{-}K^{+}\pi^{0}\gamma$\\
$ K^{+}\pi^{0}\pi^{0}\gamma$ & $ K^{+}\pi^{+}\pi^{-}\pi^{+}\pi^{-}\gamma$ & $ K^{+}\eta\pi^{-}\gamma$ & $ K^{+}K^{-}K^{+}\pi^{-}\gamma$\\
$ K_{S}\pi^{+}\pi^{-}\gamma$ & $ K_{S}\pi^{+}\pi^{-}\pi^{+}\pi^{0}\gamma$ &	$ K_{S}\eta\pi^{0}\gamma$ & $ K^{+}K^{-}K_{S}\pi^{0}\gamma$\\
$ K^{+}\pi^{-}\pi^{0}\gamma$ & $ K^{+}\pi^{+}\pi^{-}\pi^{0}\pi^{0}\gamma$ & $ K^{+}\eta\pi^{+}\pi^{-}\gamma$ &  \\
$ K_{S}\pi^{0}\pi^{0}\gamma$ & $ K_{S}\pi^{+}\pi^{-}\pi^{+}\pi^{-}\gamma$ & $ K_{S}\eta\pi^{+}\pi^{0}\gamma$ &  \\
\hline
\end{tabular}
\end{center}
\end{small}
\end{table}

\section{Angular distribution in the decays \btoksll}

The decay \btoksll\ occurs in the SM via penguin (electromagnetic penguin loop or electroweak $Z^0$ penguin loop) and $W^+W^-$ box diagrams. The effective Hamiltonian of these reactions depends in particular on the Wilson coefficients $C_i$. Particles from new physics could enter in the loop (such as charged Higgs, squarks, neutralinos, and charginos) and could lead to sizeable deviations from the SM predictions. In this section, we focus on an analysis~\cite{ref:ksll} on the angular observables in \btoksll\ (the branching fraction and rate asymmetries are addressed elsewhere in these proceedings~\cite{ref:liang}). The angular distribution depends in particular on the angle $\theta_K$, the angle between the $K$ and the $B$ in the $K^*$ rest frame, and on $\theta_\ell$, the angle between the $\ell^+$ ($\ell^-$) and the $B$ ($\Bbar$) in the $\ell^+\ell^-$ frame. From these angles, we can obtain the fraction of the longitudinal polarization of the $K^*$, $F_L$, and the lepton forward-backward asymmetry, $\mathcal{A}_{FB}$. The predicted distributions depending on these quantities, as a function of the dilepton mass squared $s=m^2_{\ell^+\ell^-}$, read:
\begin{eqnarray*}
\frac{1}{\Gamma}\frac{d\Gamma}{d\cos \theta_K} &=& \frac{3}{2} F_L \cos^2 \theta_K + \frac{3}{4}(1-F_L)(1-\cos^2 \theta_K), \\
\frac{1}{\Gamma}\frac{d\Gamma}{d\cos \theta_\ell} &=& \frac{3}{4} F_L (1 - \cos^2 \theta_\ell) + \frac{3}{8}(1-F_L)(1+\cos^2 \theta_\ell)+\mathcal{A}_{FB} \cos \theta_\ell.
\end{eqnarray*}
In the SM, at low $s$, where the effective Wilson coefficient $C_7^{\mathrm{eff}}$ dominates, $\mathcal{A}_{FB}$ is expected to be small, crossing the zero axis around $s\sim 4$~GeV$^2/c^4$. At high $s$, the product of $C_9^{\mathrm{eff}}$ and  $C_{10}^{\mathrm{eff}}$ is expected to give a large positive asymmetry. Contributions from new physics could change dramatically this picture.

We use seven bins in the $s$ variable, with the same binning as the experiments CDF, Belle, and LHCb, in order to ease the comparison and the average. Two regions in $s$ are dominated by \jpsi and \psitwos and are used as control samples (and are vetoed in the analysis). We use the decays $B^+ \to K^{*+} \ell^+ \ell^-$, followed by $K^{*+} \to K^+ \pi^0, \KS \pi^+$, and $B^0 \to K^{*0} \ell^+ \ell^-$, followed by $K^{*0} \to K^+ \pi^-$, with $\ell = e, \mu$. The decay $B^+ \to K^{*+} \mu^+ \mu^-$, $K^{*+} \to K^+ \pi^0$ is not used since it showed no improvement to the analysis. The continuum and $B\Bbar$ backgrounds are rejected using bagged decision trees (BDT), based on $\Delta E$ (the difference between the expected $B$ energy and the reconstructed $B$ energy), event shape and vertexing variables. A likelihood ratio $R$ is constructed from the $B\Bbar$ BDT. The angular observables are extracted from simultaneous fits over the combinations of final states. The strategy in each bin in $s$ is three-fold: first, for each of the five decay modes, we fit the variables \mes, $M(K\pi)$ (the reconstructed invariant mass of the $K^*$), and $R$, where we fix the fit parameters for the next stage; second, we fit the $\cos \theta_K$ distribution to extract $F_L$; and third, we fix $F_L$, and fit the $\cos \theta_\ell$ distribution to extract $\mathcal{A}_{FB}$.

\begin{figure}[htb]
\begin{center}
\includegraphics[width=7.5cm]{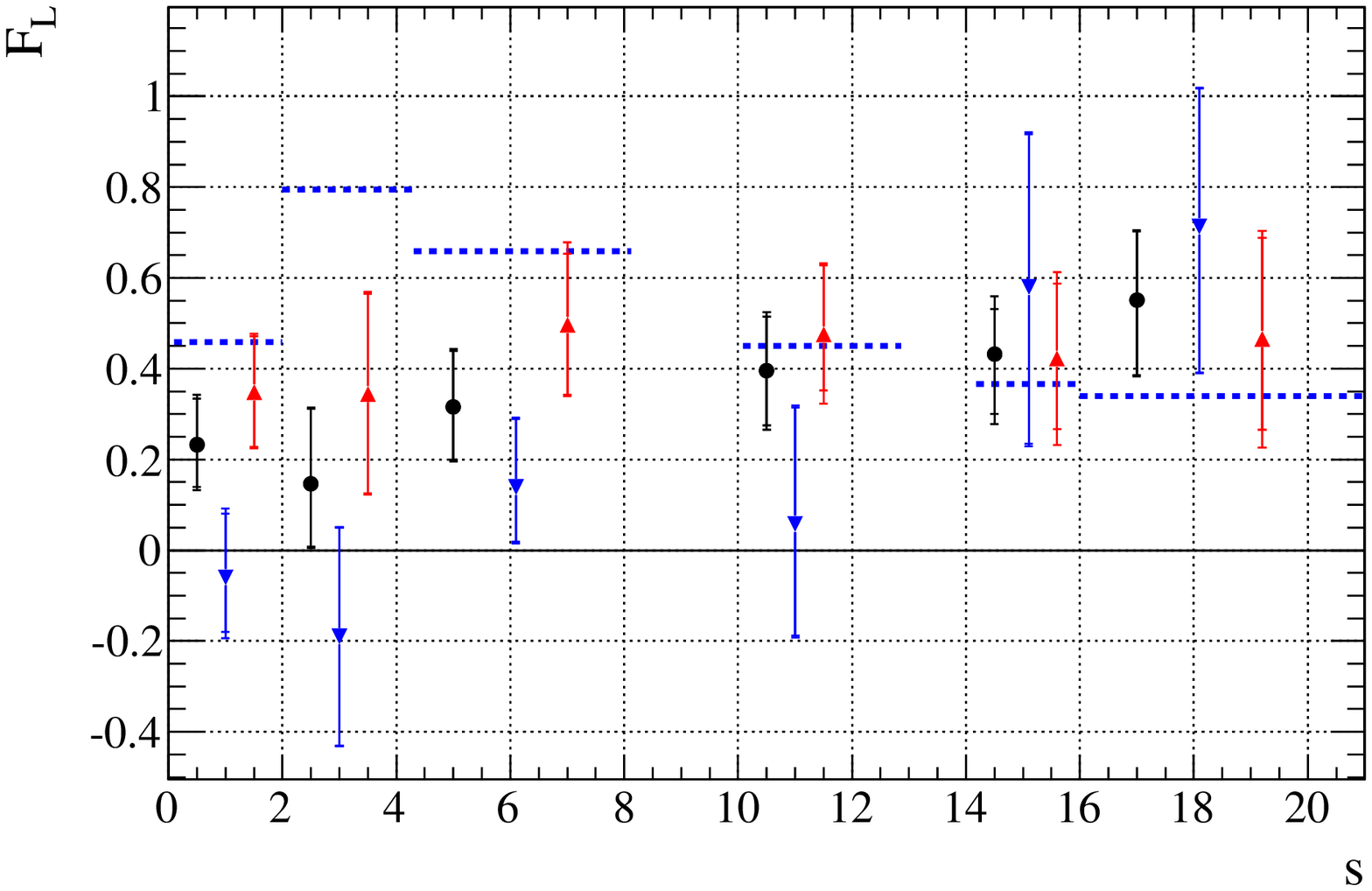}
\includegraphics[width=7.5cm]{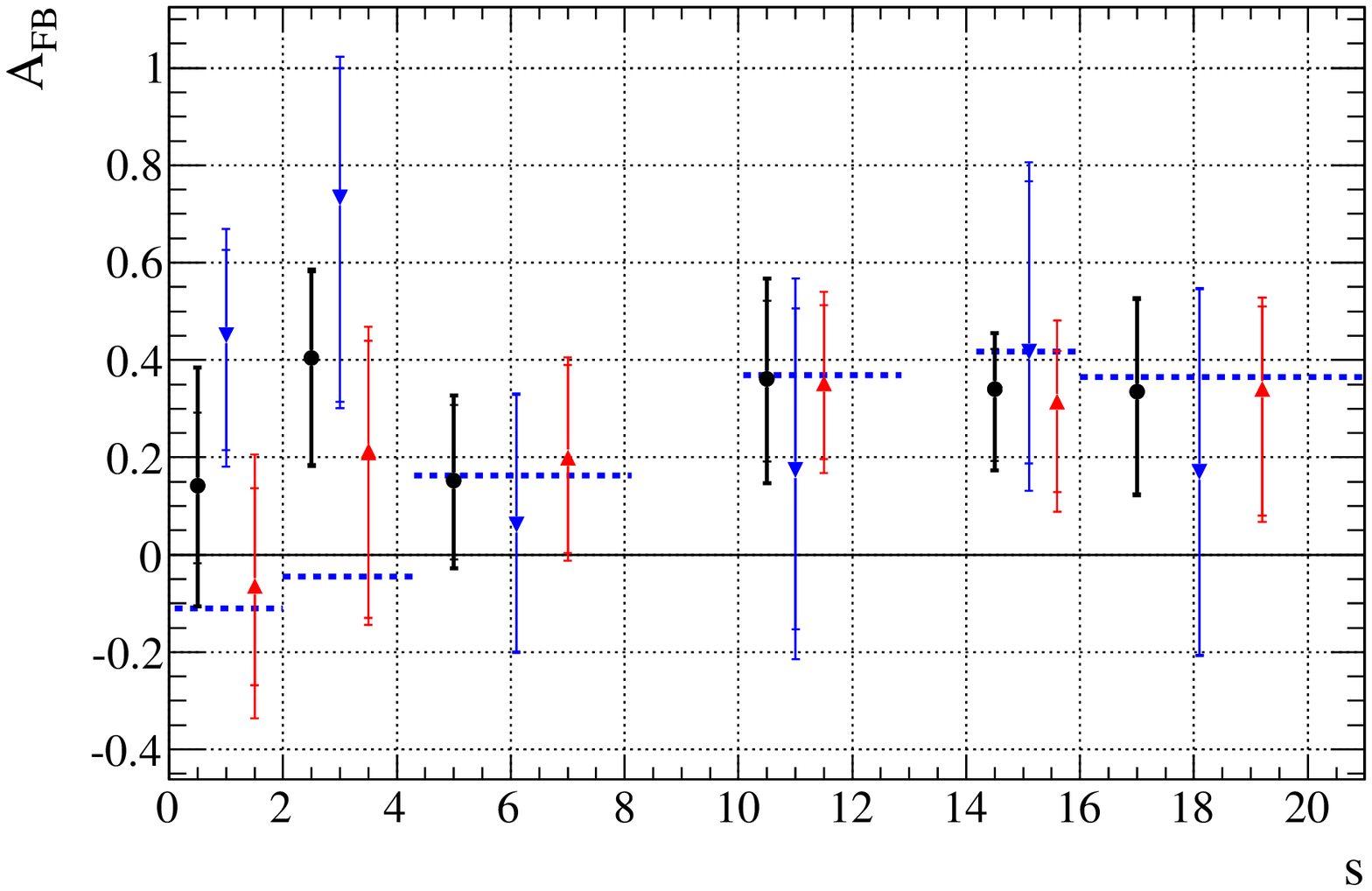}
\caption{Distributions of $F_L$ (left) and $\mathcal{A}_{FB}$ (right) in the seven bins in the $s=m^2_{\ell^+\ell^-}$ variable. The two vetoed regions correspond to regions dominated by the \jpsi and \psitwos resonances. The blue downward triangles correspond to $B^+ \to K^{*+}\ell^+\ell^-$, the red upward triangles to $B^0 \to K^{*0}\ell^+\ell^-$, and the black dots to the combination $B \to K^{*}\ell^+\ell^-$. The blue dotted line shows the SM prediction (the theoretical uncertainties are of the order of 5-10\% at low $s$ and 10-15\% at high $s$).}
\label{fig:ksll}
\end{center}
\end{figure}

The results are presented in Fig.~\ref{fig:ksll}. We obtain significantly more precise values than the ones from either Belle~\cite{ref:belleksll} or CDF~\cite{ref:cdfksll}. Other experiments, especially LHCb~\cite{ref:lhcbksll}, are dominated by the $B^0 \to K^{*0}\ell^+\ell^-$ channels. On the contrary, we are able to study $B^+ \to K^{*+}\ell^+\ell^-$ and notice some tension at low $s$ in this channel.

\section{Search for lepton-number violating processes in $B^+ \to h^- \ell^+ \ell^+$ decays}

\begin{figure}[htb]
\begin{center}
\includegraphics[width=5.5cm]{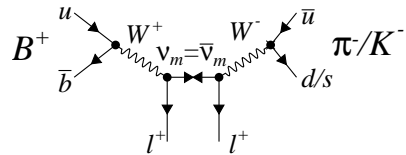}
\caption{An example of a diagram of the process $B^+ \to h^- \ell^+ \ell^+$ with lepton-number violation.}
\label{fig:diagMaj}
\end{center}
\end{figure}

In the SM, the lepton number $L$ is conserved in low-energy collisions and decays. However, since the neutrinos are oscillating, we know that these particles have a mass. If the neutrinos are of Majorana type, then $L$ violation becomes possible. This violation could be seen in $B$ meson decays via the process $B^+ \to h^- \ell^+ \ell^+$ (see 
Fig.~\ref{fig:diagMaj}). This process becomes resonant if the neutrino mass lies between the $h$ meson and the $B$ meson masses. Such processes involving meson decays~\cite{ref:lvtheo} are an alternative to neutrinoless double beta decays. Beyond the SM, this diagram could lead to a lepton violation via new physics, such as left-right symmetric gauge theories, SO(10) supersymmetry, $R$-parity violating models, or extra-dimensions.

\begin{figure}[htb]
\begin{center}
\includegraphics[width=7.5cm]{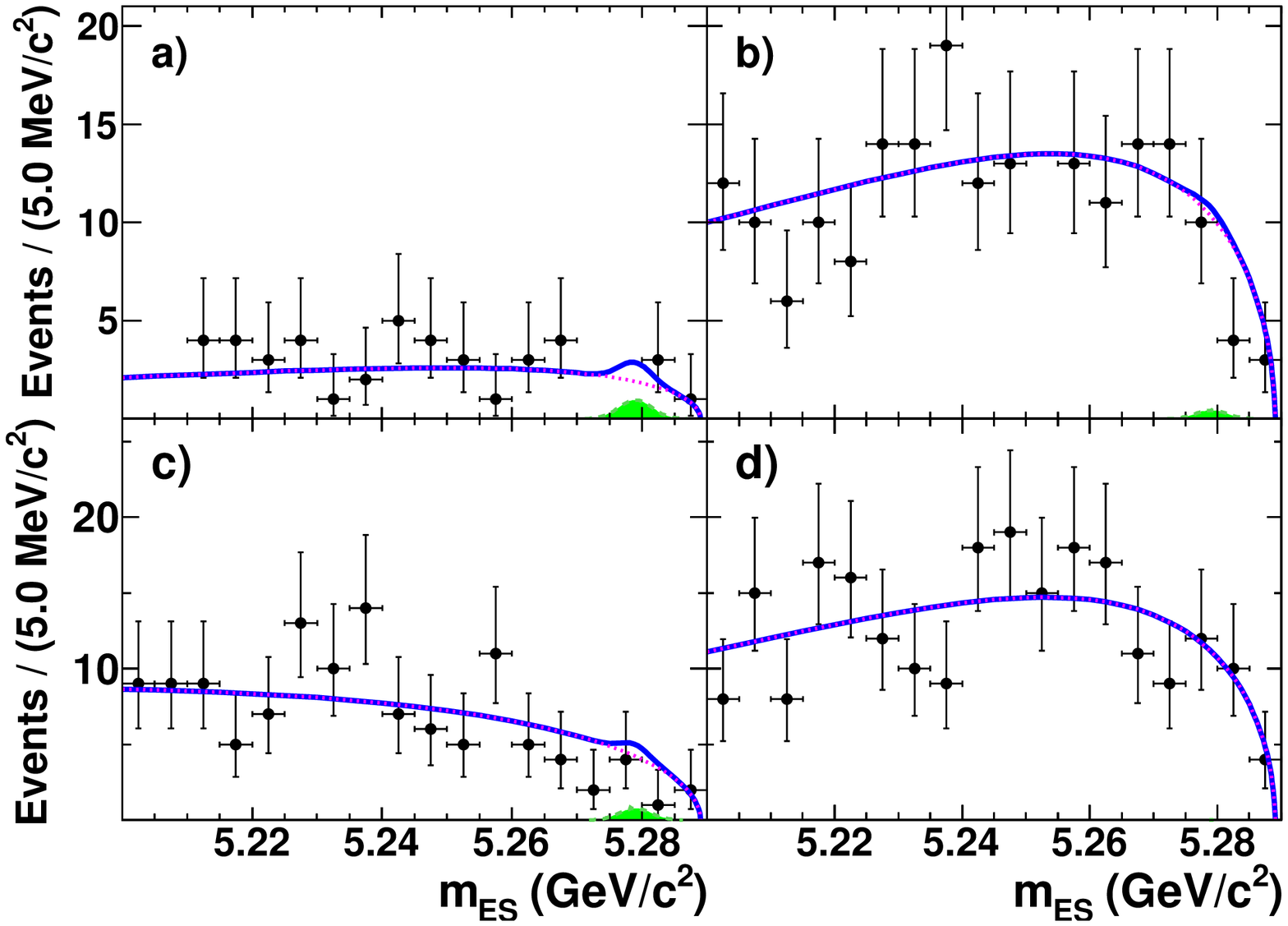}
\hspace{0.7cm}
\includegraphics[width=7.5cm]{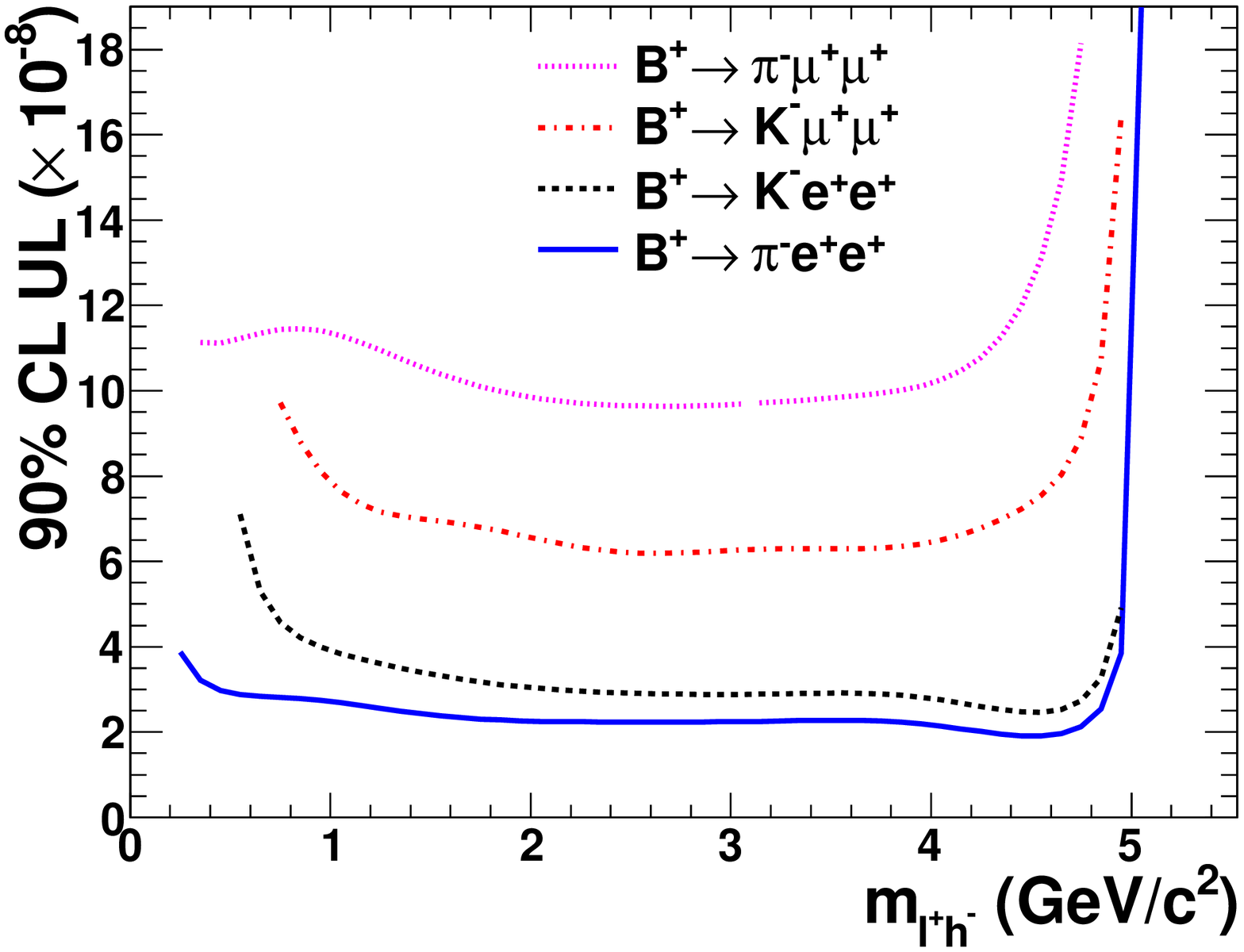}
\caption{Left: \mes\ distributions for a) $B^+ \to K^- e^+ e^+$, b) $B^+ \to K^- \mu^+ \mu^+$, c) $B^+ \to \pi^- e^+ e^+$, and d) $B^+ \to \pi^- \mu^+ \mu^+$.
The solid blue line is the total fit, the dotted magenta line is
the background, the solid green histogram is the signal. Right: 90\% CL upper limits on the branching fractions as a function of the mass $m_{\ell^+ h^-}$.}
\label{fig:hll}
\end{center}
\end{figure}

In this analysis~\cite{ref:lv}, four final states are considered: the decay channel is $B^+ \to h^- \ell^+ \ell^+$ with $h=K,\pi$ and $l=e,\mu$. The selection is identical to the previous section since the final states are very similar. Unbinned maximum likelihood fits of \mes\ and $R$, the likelihood ratio, are performed for each of the four modes. We use the $B^+ \to \jpsi h^+$ data control sample to obtain the \mes\ fit parameters. The fits are shown in Fig.~\ref{fig:hll}, where we observe that no signal is seen. From these negative results, we obtain upper limits at 90\% confidence level (CL) on the four channels: $\mathcal{B}(B^+ \to K^- e^+ e^+)<3.0\times10^{-8}$, $\mathcal{B}(B^+ \to K^- \mu^+ \mu^+)<6.7\times10^{-8}$, $\mathcal{B}(B^+ \to \pi^- e^+ e^+)<2.3\times10^{-8}$, and $\mathcal{B}(B^+ \to \pi^- \mu^+ \mu^+)<10.7\times10^{-8}$. These upper limits are 40-70 times more stringent than previous limits set by other experiments. These results can be translated on upper limits on the branching fractions as a function of the mass $m_{\ell^+ h^-}$ (Fig.~\ref{fig:hll}), which can be related to the Majorana neutrino mass for diagrams of the type of Fig.~\ref{fig:diagMaj}.

\section{A search for the decay modes $B^\pm \to h^\pm \tau \ell$}

FCNC and charged lepton flavor violation are forbidden in the SM at tree level. However, in many extensions of the SM, these effects could be enhanced, especially for the second and third generation~\cite{ref:htltheo}. We study~\cite{ref:htl} the eight final states $B^\pm \to h^\pm \tau \ell$, with $h=K,\pi$ and $\ell = e, \mu$. The final states $B^\pm \to K^\pm \tau e$, $B^\pm \to \pi^\pm \tau \mu$, and $B^\pm \to \pi^\pm \tau e$ have never been done before. 
We fully reconstruct the hadronic $B$ on one side (the ``tag'' $B$) using final states of the type $B^- \to D^{(*)0}X^-$, where $X^-$ is composed of $\pi^\pm$, $K^\pm$, \KS, and $\pi^0$. This determines the three-momentum of the other $B$ (the ``signal'' $B$) on the other side and thus allows us to indirectly reconstruct the $\tau$ lepton through:
\begin{eqnarray*}
\vec p_\tau & = & -\vec p_{\rm tag} - \vec p_h - \vec p_\ell, \\
E_\tau & = & E_{\rm beam} - E_h - E_\ell, \\
m_\tau & = & \sqrt{ E_\tau^2 - |\vec{p}_\tau|^2 },
\end{eqnarray*}
where ($E_\tau$, $\vec p_\tau$), ($E_h$, $\vec p_h$), and ($E_\ell$, $\vec p_\ell$)
are the corresponding four-momenta of the reconstructed signal objects, and where $\vec p_{\rm tag}$ is the three-momentum of the tag $B$, and $E_{\rm beam}$ the beam energy. The $\tau$ is required to decay to a ``one-prong'' final state: $\tautoenunu$, $\tautomununu$, and $\tautopinu$ with $n \ge 0$. The signal branching fraction is determined by using the ratio of the number of $B^\pm \to h^\pm \tau \ell$ signal candidates to the yield of control samples of $B^+ \to \Dbar^{(*)0} \ell^+ \nu; \Dbar^0 \to K^+ \pi^-$ events from a fully reconstructed hadronic $B^\pm$ decay sample.

The background is mainly coming from semileptonic $B$ decays when the charge of the signal $B$ is the same as the one from the primary lepton, and mainly from semileptonic $D$ decays when the charges are opposite. We remove these backgrounds by rejecting the signal $B$ candidates where two of their daughters are kinematically compatible with originating from a charm decay. After this requirement, we reject the continuum background using a cut on the likelihood ratio $R$, based on particle identification and event shape variables.

\begin{figure}
     \begin{center}
     \includegraphics[width=0.4\linewidth]{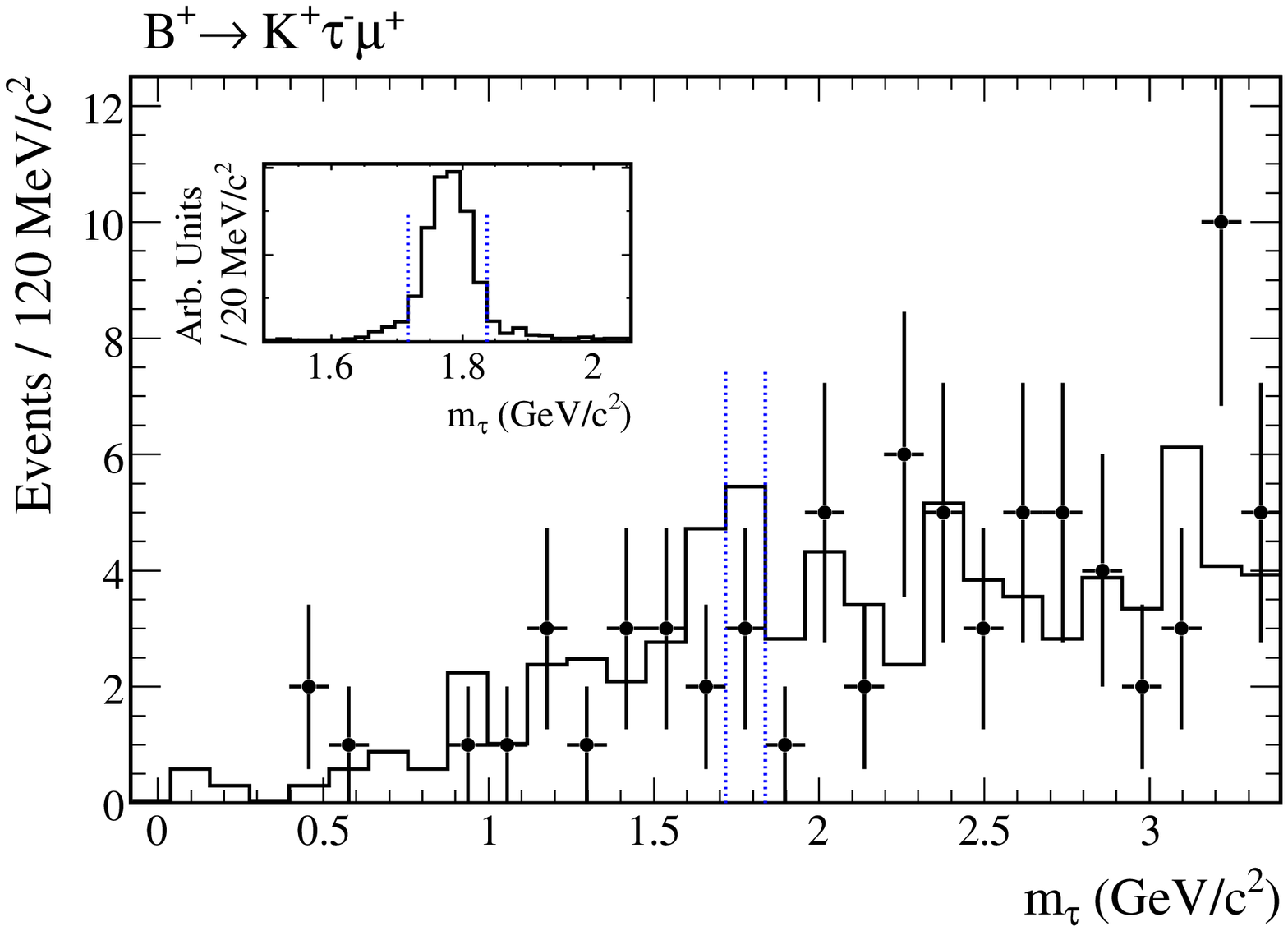}
     \includegraphics[width=0.4\linewidth]{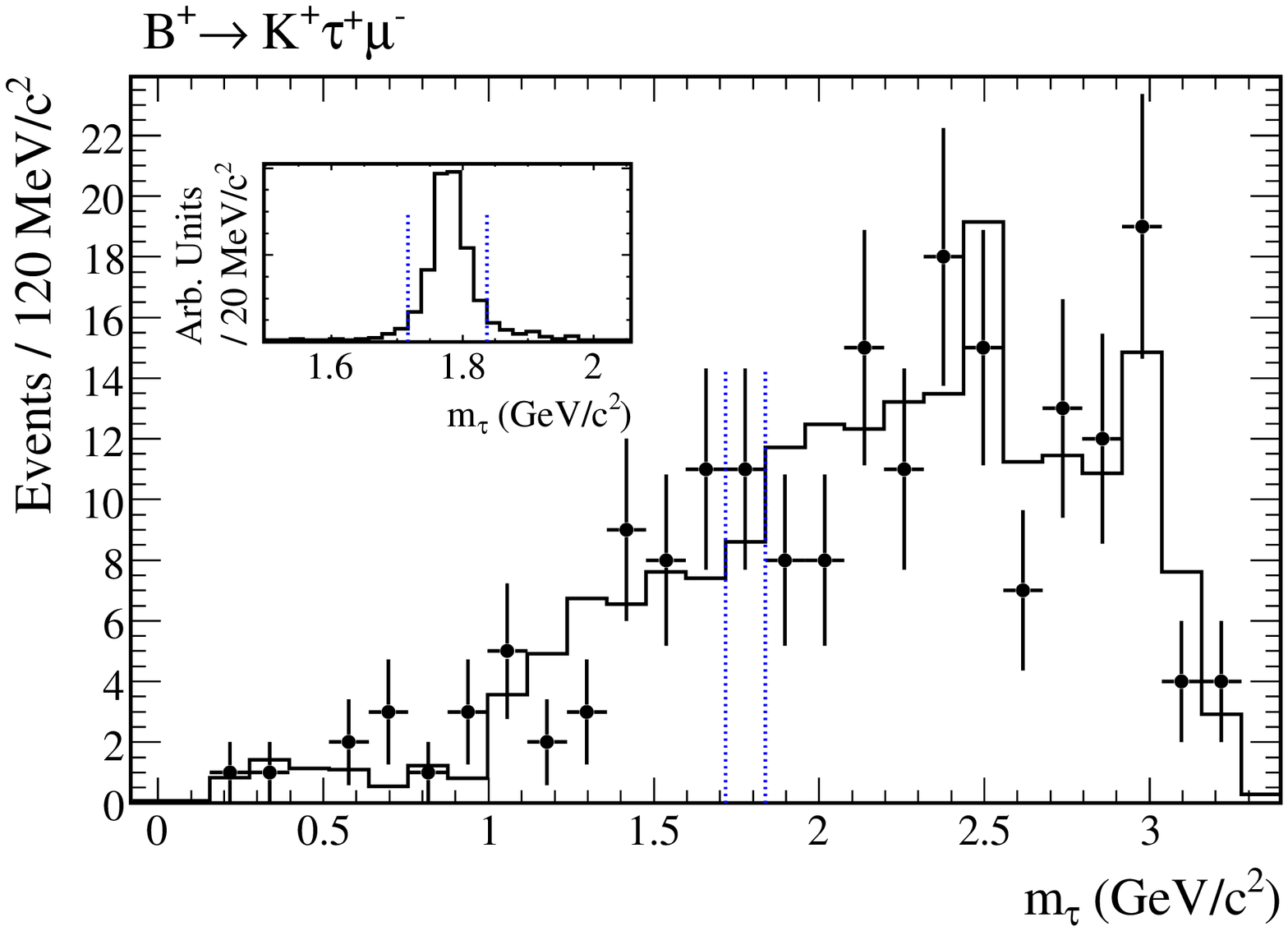}
     \includegraphics[width=0.4\linewidth]{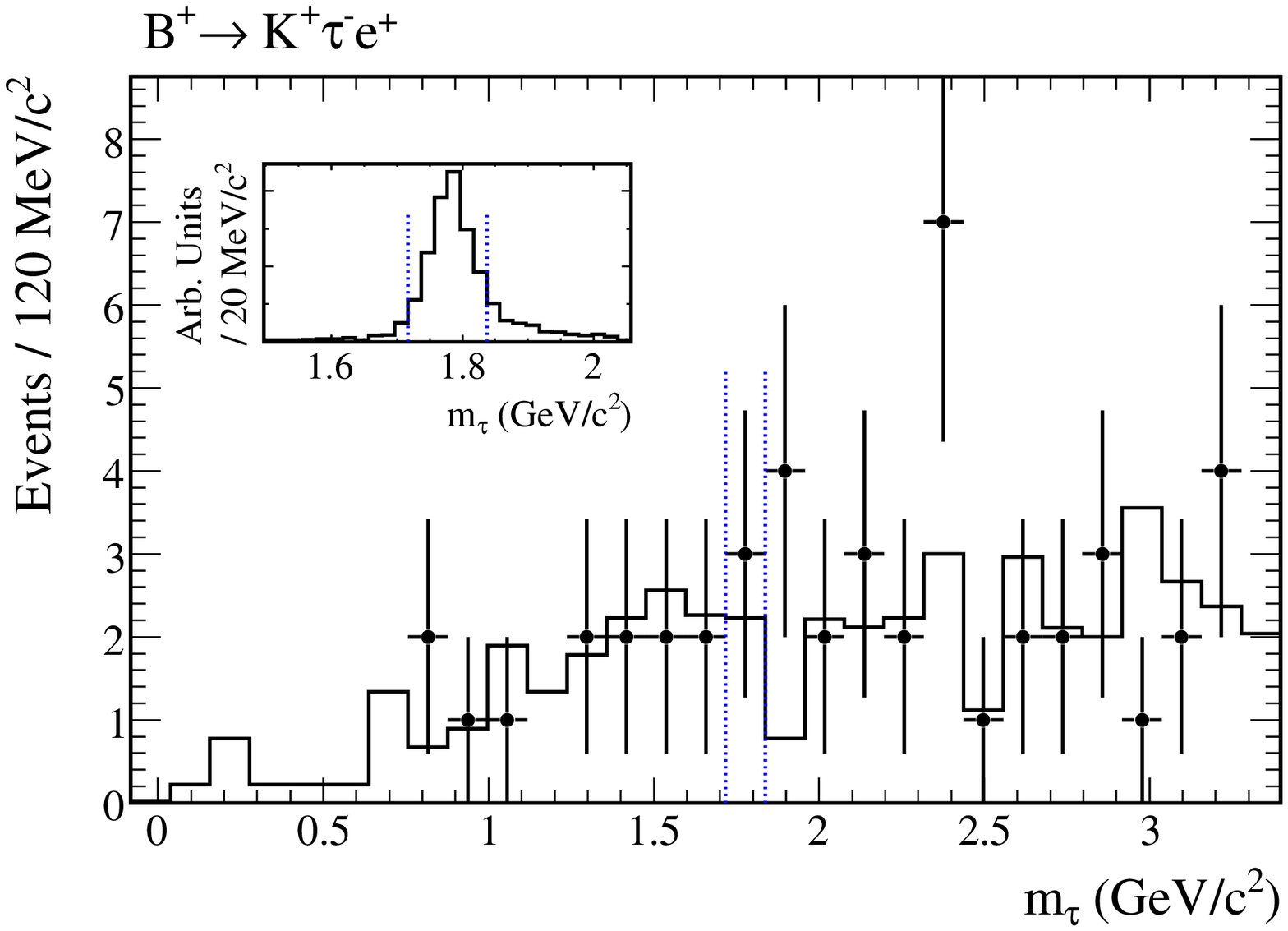}
     \includegraphics[width=0.4\linewidth]{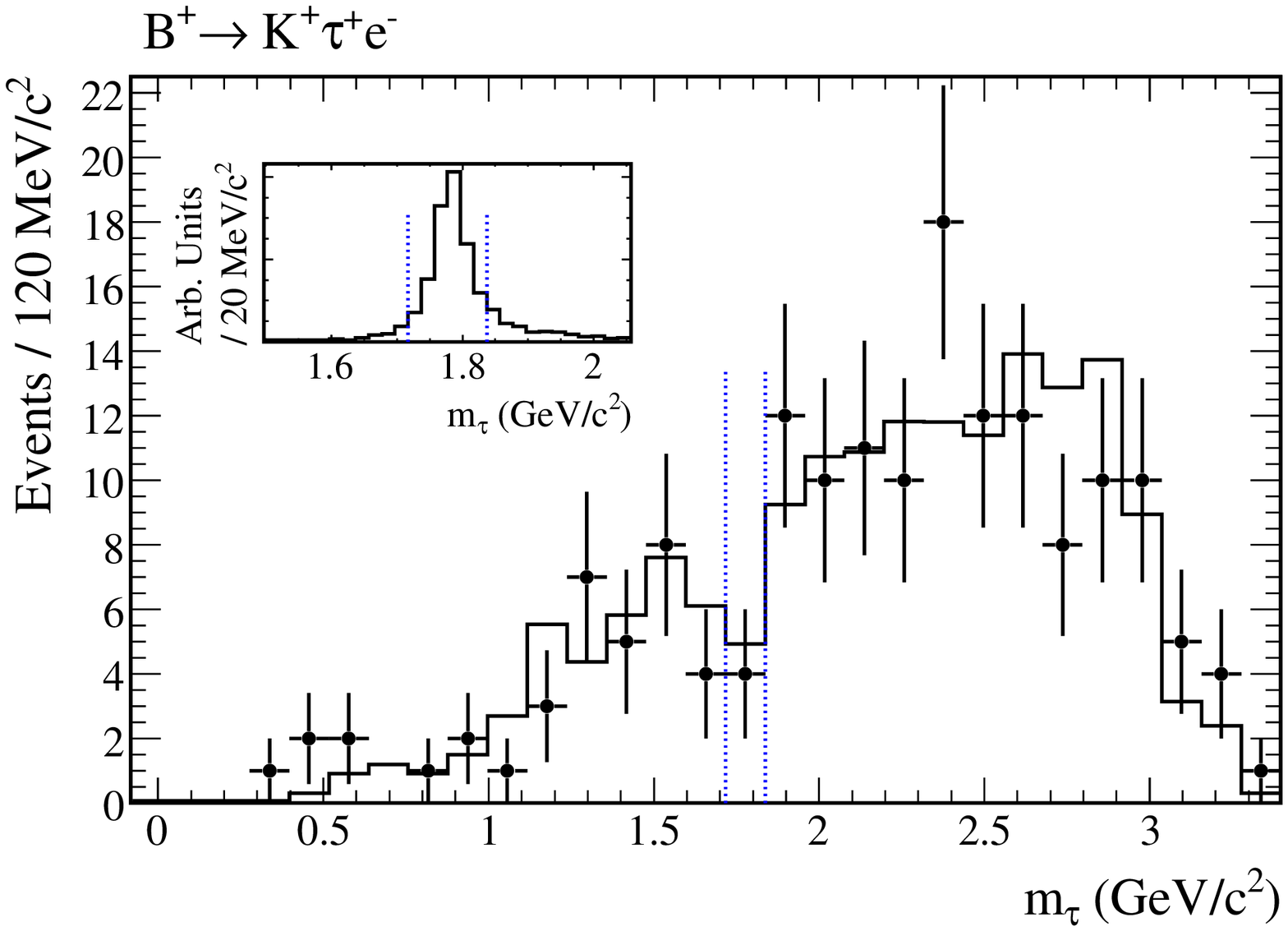}
     \caption{
         Observed distributions of the $\tau$ invariant mass for the \btoktl\ modes.
         The distributions show the sum of the three $\tau$ channels ($e$, $\mu$, $\pi$).
         The points with error bars are the data.
         The solid line is the background MC which has been normalized to the
         area of the data distribution.
         The dashed vertical lines indicate the $m_\tau$ signal window range.
         The inset shows the $m_\tau$ distribution for signal MC.
     }
     \label{fig:mtau-ktaul}
     \end{center}
   \end{figure}

\begin{figure}
     \begin{center}
     \includegraphics[width=0.4\linewidth]{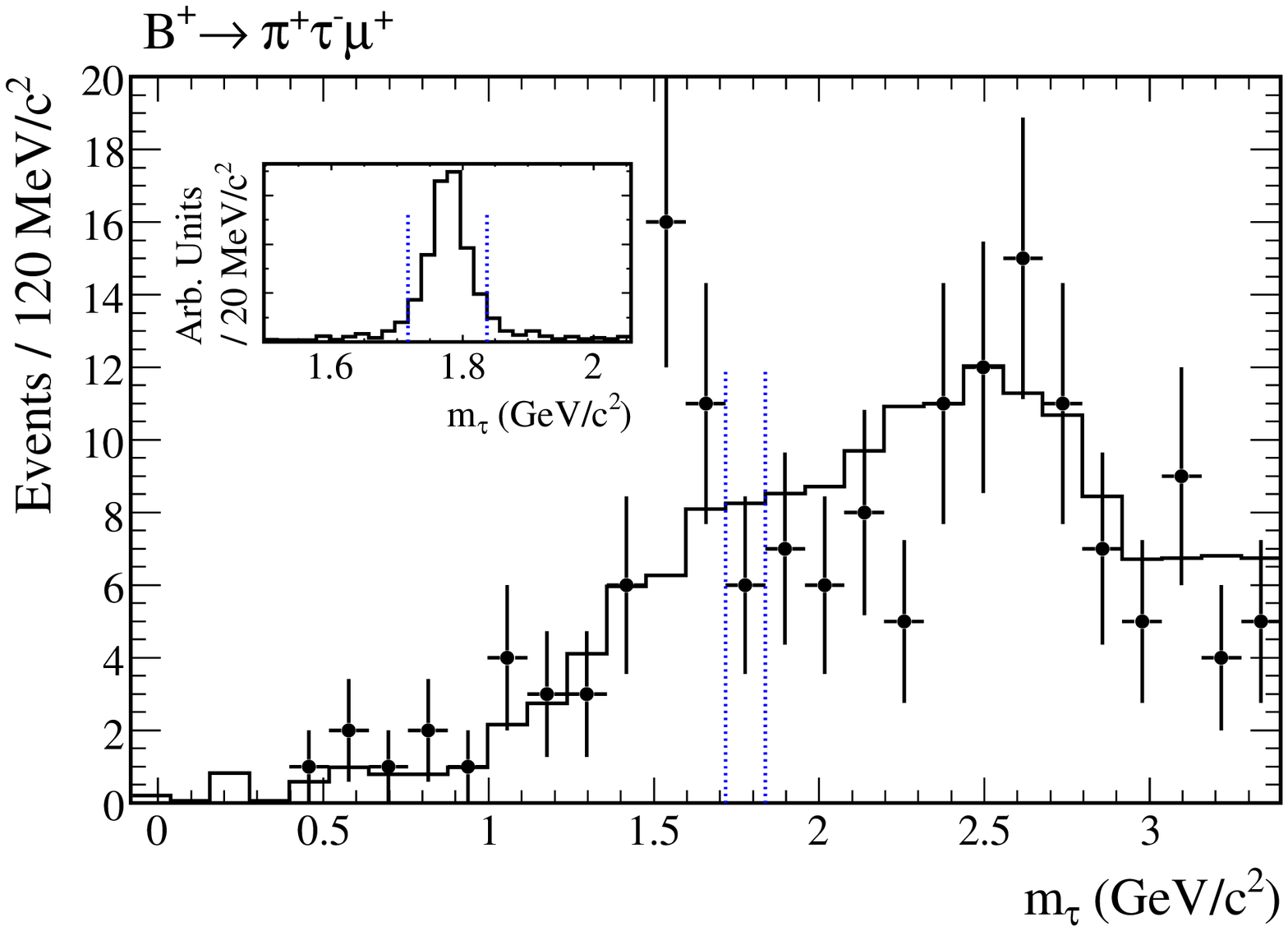}
     \includegraphics[width=0.4\linewidth]{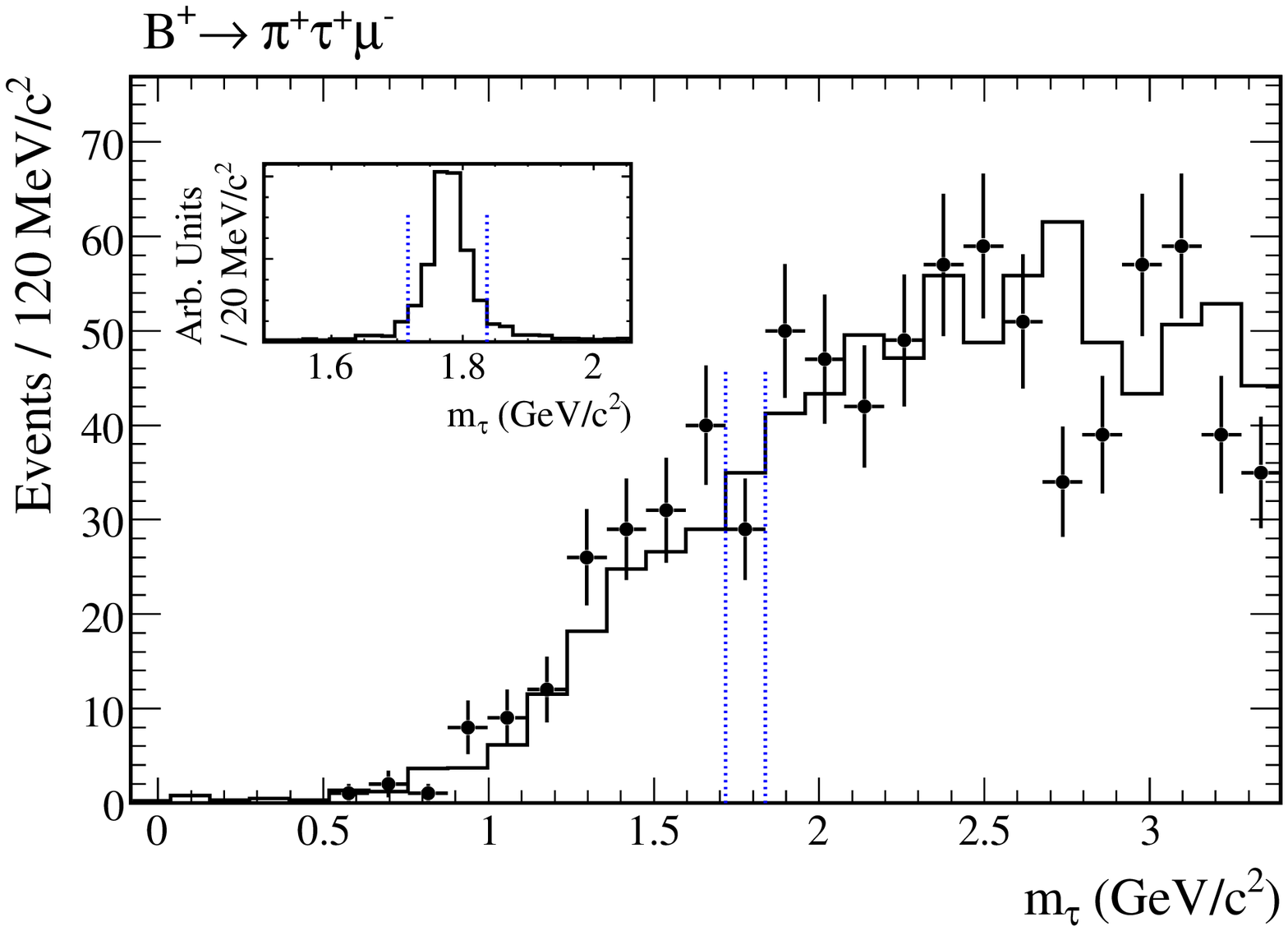}
     \includegraphics[width=0.4\linewidth]{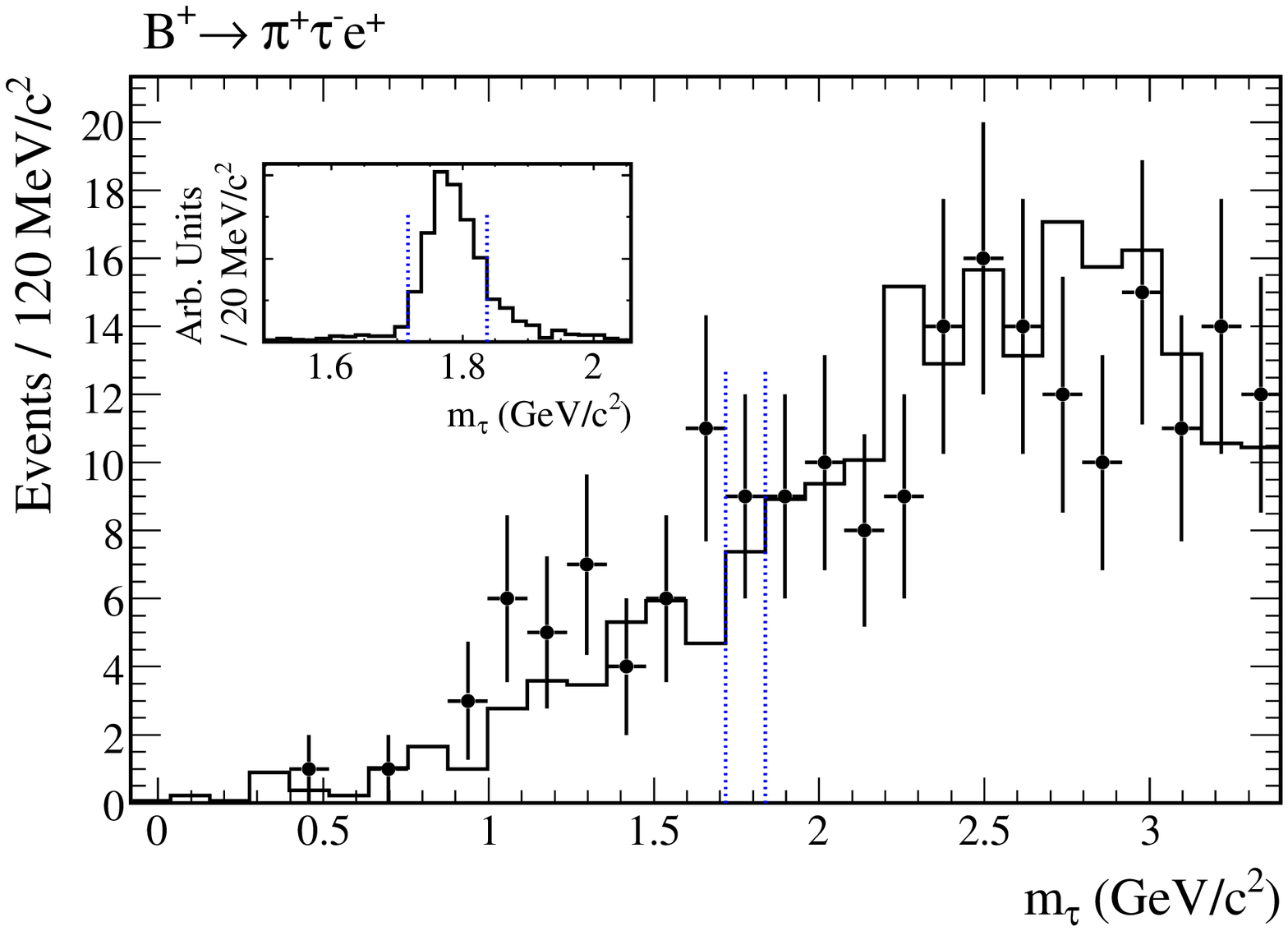}
     \includegraphics[width=0.4\linewidth]{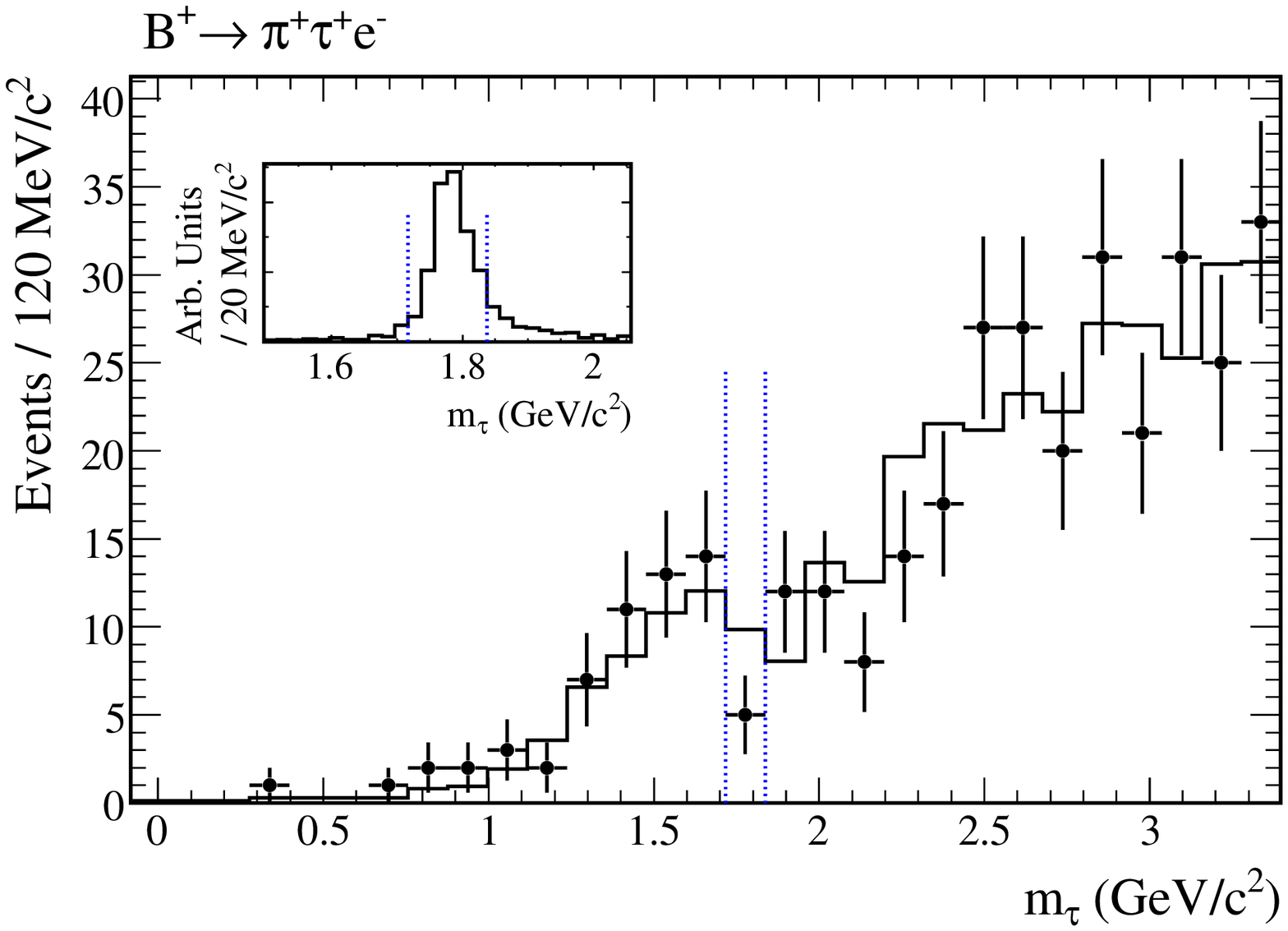}
     \caption{
         Observed distributions of the $\tau$ invariant mass for the \btoptl\ modes.
         The distributions show the sum of the three $\tau$ channels ($e$, $\mu$, $\pi$).
         The points with error bars are the data.
         The solid line is the background MC which has been normalized to the
         area of the data distribution.
         The dashed vertical lines indicate the $m_\tau$ signal window range.
         The inset shows the $m_\tau$ distribution for signal MC.
     }
     \label{fig:mtau-pitaul}
     \end{center}
   \end{figure}

The signal region is defined as $\pm 60 \mevcc$ around the indirectly reconstructed $\tau$ mass $m_\tau$. Figures~\ref{fig:mtau-ktaul} and \ref{fig:mtau-pitaul} show the $m_\tau$ distributions for the data, for the background, and for the signal MC. No signal is observed, which allows us to put 90\% CL limit on the branching fractions. Assuming ${\cal B}(B^+ \to h^+ \tau^- \ell^+) = {\cal B}(B^+ \to h^+ \tau^+ \ell^-)$, we obtain the combined limits shown in Table~\ref{tab:combined-6chan-limits}. These limits can be translated into model-independent bounds on the energy scale of new physics in flavor-changing operators~\cite{ref:htlscale}: $\Lambda_{\bar b d}>11$~TeV and $\Lambda_{\bar b s}>15$~TeV (at 90\% CL), which improved the previous limits of 2.2 and 2.6~TeV, respectively.

   \begin{table}
     \begin{center}
       \caption{
          Branching fraction central values and 90\% CL upper limits (UL)
          for the combination
          ${\cal B}(B^+ \to h^+ \tau \ell) \equiv {\cal B}(B^+ \to h^+ \tau^- \ell^+) + {\cal B}(B^+ \to h^+ \tau^+ \ell^-)$.
       }
       \label{tab:combined-6chan-limits}
       \begin{tabular}{|l|c|c|}
         \hline
                      &  \multicolumn{2}{|c|}{ ${\cal B}(B \to h \tau \ell)$ $(\times 10^{-5})$} \\
          \ \ \ \ \  Mode \ \ \ \ \ \ \ \ \ \ \        &  \multicolumn{2}{|c|}{\ \ \ \ central value\ \ \ \  \ \ \ 90\% CL UL \ \ \ } \\
         \hline
         $B^+ \to K^+ \tau \mu$   &  ~~~~~~$0.0\ ^{+2.7}_{-1.4}~~~~~~$   &  $<4.8$   \\
         \hline
         $B^+ \to K^+ \tau e$     &  $-0.6\ ^{+1.7}_{-1.4}$  &  $<3.0$   \\
         \hline
         $B^+ \to \pi^+ \tau \mu$ &  $0.5\ ^{+3.8}_{-3.2}$  &  $<7.2$   \\
         \hline
         $B^+ \to \pi^+ \tau e$   &  $2.3\ ^{+2.8}_{-1.7}$  &  $<7.5$   \\
         \hline
       \end{tabular}
     \end{center}
   \end{table}

\section{Search for $CP$ violation in the decays $\tau^-\to\pi^-\KS\left(\geq 0\pi^0\right)\nut$}

$CP$ violation, until now, has only been observed in hadronic decays ($K$, $B$, and $D$ systems). However, Bigi and Sanda predict~\cite{ref:bigi} a non-zero decay rate asymmetry for $\tau$ decays to final states containing a $\KS$ meson, due to the $CP$ violation in the kaon sector. The decay rate asymmetry is:
\begin{equation*}
{\asy} = \frac{\Gamma\left({\taupksb}\right) - \Gamma\left({\taupks}\right)}
              {\Gamma\left({\taupksb}\right) + \Gamma\left({\taupks}\right)}, 
\end{equation*}
and is predicted to be equal to $(0.33\pm0.01)\%$. Any deviation from the SM prediction could be a sign of new physics. It has to be noted that $\asy$ is independent of the number of neutral pions in the final state. Recently, Grossman and Nir~\cite{ref:grossman} noticed that the calculation needs to take into account interferences between the amplitudes of intermediate \KS and \KL mesons, which are as important as the pure \KS amplitude. This means that $\asy$ depends on the reconstruction efficiency as a function of the $\KS \rightarrow \pi^+\pi^-$ decay time.

We study here~\cite{ref:cpv} the decay channel $\tau^-\to\pi^-\KS\left(\geq 0\pi^0\right)\nut$. The event is divided into two hemispheres, one corresponding to the signal side, and one to the tag side with $\tau^- \to \ell^- \overline{\nu}_\ell \nut$, $\ell=e,\mu$. The selection of the signal events requires that the invariant mass of the reconstructed $\tau$ lepton is smaller than $1.8 \gevcc$. A first likelihood ratio is constructed to reject continuum background based on energy, calorimeter clusters, thrust, and momentum. A second likelihood ratio is aimed at reducing the \KS background based on \KS reconstruction parameters. After the selection, we obtain $199064$ candidates for the electron tag channel ($e$-tag), and $140602$ for the muon tag channel ($\mu$-tag). The composition of the sample in term of signal and background events is presented in Table~\ref{tab:cpv}.

\begin{table}[ht]
\caption{Composition of the sample after all selection criteria have been applied.}
\label{tab:cpv}
\begin{center}
\begin{tabular}{|l|c|c|}
\hline
Source                         & \multicolumn{2}{|c|}{Fractions (\%) } \\
                               & \multicolumn{2}{|c|}{\hspace{0.5cm} $e$-tag \hspace{0.5cm} \hspace{0.5cm} $\mu$-tag \hspace{0.5cm}} \\
\hline
$\tau^- \rightarrow \pi^- \,\KS (\geq 0\pi^0) \, \nut$  & $78.7 \pm 4.0$  & $78.4 \pm 4.0$  \\
$\tau^- \rightarrow K^-  \,\KS (\geq 0\pi^0) \, \nut$  & $4.2 \pm 0.3$   & $4.1 \pm 0.3$   \\
$\tau^- \rightarrow \pi^- \,K^0 \overline{K}^0 \, \nut$          & $15.7 \pm 3.7$  & $15.9 \pm 3.7$  \\
Other background      & $1.40 \pm 0.06$ & $1.55 \pm 0.07$  \\
\hline
\end{tabular}
\end{center}
\end{table}

We need to correct the raw asymmetry from the pollution of the other modes shown in Table~\ref{tab:cpv}. For the mode $\tau^- \rightarrow K^-  \,\KS (\geq 0\pi^0) \, \nut$, the expected asymmetry is opposite to the one from the signal, and for the mode $\tau^- \rightarrow \pi^- \,K^0 \overline{K}^0 \, \nut$, the expected asymmetry is zero. Furthermore, an additional correction was pointed out recently~\cite{ref:ko}: we need to take into account a correction on the asymmetry $\asy$ due to the different nuclear-interaction cross-section of the $K^0$ and $\overline{K}^0$ mesons with the material in the detector. We calculate this correction to be $(0.07\pm0.01)\%$ and we subtract it from the measured asymmetry. After all corrections are applied, and after combining the results from the $e$-tag and $\mu$-tag, we obtain $\asy=(-0.36\pm 0.23\pm 0.11)\%$.

As we mentioned, this result should be compared with the prediction of the SM, corrected by the $\KS \rightarrow \pi^+\pi^-$ decay time dependence. Using the MC sample, we find a multiplicative factor of $1.08\pm0.01$. The SM decay-rate asymmetry, after correction, is then predicted to be $\asy^{\mathrm{SM}}=(0.36\pm0.01)\%$. We observe that our measurement is 2.8 standard deviations away from the SM prediction.

\section{Conclusions}

The \babar\ collaboration, four years after its shutdown, is still producing competitive results, with many more to come in the near future. We have presented here five new results. In general, we have a good agreement with the prediction of the SM, except for two studies: we observe some tensions at low $m^2_{\ell^+\ell^-}$ for $F_L$ and $\mathcal{A}_{FB}$ in the channel $B^+ \to K^{*+}\ell^+\ell^-$; and we measure a $CP$ violation parameter 2.8 standard deviations away from the SM prediction in the channel $\tau^-\to\pi^-\KS\left(\geq 0\pi^0\right)\nut$. The actual statistics is not sufficient to tell whether or not these could be indication for new physics.

\end{document}